\documentclass[a4paper,fleqn]{cas-sc}

\usepackage[authoryear,longnamesfirst]{natbib}
\usepackage{algorithm}

\newcommand{\bV}{\mathbf{V}}

\newcommand{\bW}{\mathbf{W}}

\newcommand{\bM}{\mathbf{M}}

\newcommand{\balpha}{\boldsymbol{\alpha}}
\newcommand{\bm}[1]{\mbox{\boldmath$ #1 $\unboldmath}}

\newcommand{\beq}{\begin{equation}}
\newcommand{\eeq}{\end{equation}}

\newcommand{\be}{\begin{eqnarray}}
\newcommand{\ee}{\end{eqnarray}}

\newcommand{\bes}{\begin{eqnarray*}}
\newcommand{\ees}{\end{eqnarray*}}

\newcommand{\bI}{\mathbf{I}}

\newcommand{\bbeta}{\boldsymbol{\beta}}
\newcommand{\btheta}{\boldsymbol{\theta}}

\newcommand{\bS}{\mathbf{S}}

\def\bSig{\mbox{\boldmath $\Sigma$}}

\def\bTheta{\mbox{\boldmath $\Theta$}}
\def\btheta{\mbox{\boldmath $\theta$}}

\newtheorem{proposition}{Proposition}

\newdefinition{remark}{Remark}
\newproof{proof}{Proof}

\begin{document}
\let\WriteBookmarks\relax
\def\floatpagepagefraction{1}
\def\textpagefraction{.001}
\shorttitle{Unifying precision and covariance estimation}
\shortauthors{Dallakyan and Pourahmadi}

\title[mode=title]{Unifying Sparse Estimation of Precision and Covariance Matrices}

\author[1]{Aramayis Dallakyan}[orcid=0000-0002-3897-8261]
\cormark[1]
\ead{dallakyan.aramayis@gmail.com}
\credit{Conceptualization of this study, Methodology, Software}
\affiliation[1]{organization={StataCorp LLC},
                city={College Station},
                postcode={77845},
                state={Texas},
                country={USA}}

\author[2]{Mohsen Pourahmadi}
\credit{Methodology, Writing}
\affiliation[2]{organization={Department of Statistics, Texas A\&M University},
                city={College Station},
                postcode={77845},
                state={Texas},
                country={USA}}

\cortext[cor1]{Corresponding author}

\begin{abstract}
Penalized maximum likelihood is a widely used approach to estimate a sparse  precision or a sparse covariance matrix. A zero in the 
precision matrix encodes a conditional
 independence and a zero in the covariance matrix a marginal independence, and the two estimators have been developed separately, each
 with its own algorithm. We show that these two problems can
 be unified via a single reparametrization of the last column of either matrix.
 This reparametrization decouples the likelihood function for each problem
 to a one-dimensional minimization followed by a penalized multiple regression. Our approach applies to any penalty that is separable in the off-diagonal
 entries and linear on the diagonal. When the penalty is the $\ell_1$ norm,
 it  unifies the graphical Lasso and the covariance graphical Lasso
 algorithms.
\end{abstract}


\maketitle

\section{Introduction}

\noindent
Covariance and precision matrices capture various aspects of  dependence structure among random variables of interest. Under a multivariate Gaussian assumption, a zero entry in the covariance matrix implies that the corresponding pair of variables is marginally independent, whereas a zero entry in the precision matrix implies that the pair is conditionally independent given the remaining variables. Sparse estimation of these matrices in high-dimensional settings is not only essential for revealing the underlying relationships among variables but also fundamental to a wide range of multivariate procedures, such as principal component analysis, discriminant analysis, and regression \citep{Anderson:2003, pourahmadi2026}.

Let $Y_1,\ldots,Y_n$ be a mean-zero sample from $N_p(0,\bSig)$ with sample covariance $\bS$. Consider estimating a target matrix $\bM$, which is either the precision matrix $\bTheta=\bSig^{-1}$ or the covariance matrix $\bSig$, by solving the optimization problem
\begin{equation}\label{e:general}
\min_{\bM\succ 0}\; \ell(\bM)+\tau\,\mbox{tr}(\bM)+\sum_{j\neq k}P_\lambda(|m_{jk}|),
\end{equation}
where the loss is the Gaussian negative log-likelihood in the chosen parametrization,
\begin{equation}\label{e:loss}
\ell(\bM)=\begin{cases}
-\log|\bM|+\mbox{tr}(\bS\bM), & \bM=\bTheta,\\[2pt]
\log|\bM|+\mbox{tr}(\bS\bM^{-1}), & \bM=\bSig,
\end{cases}
\end{equation}
and $|\cdot|$ is the determinant.  
Here $P_\lambda$ penalizes the off-diagonal entries and $\tau\ge 0$ penalizes the diagonal.
Setting $\tau=0$ leaves the diagonal unpenalized, which is the version most often used in practice.
In Proposition~\ref{p:affine}, we show that the linearity of a diagonal penalty  $\tau\mbox{tr}(\bM)$ cannot be relaxed and is required for our approach.
 We assume $P_\lambda$ is continuous on $[0,\infty)$ with $P_\lambda(0)=0$. 

\noindent
\textbf{The $\ell_1$ penalty.} The general problem (\ref{e:general}) subsumes some of the well-known
approaches of precision and covariance matrix estimation. 
Let the penalty be defined as $P_\lambda(t)=\lambda |t|$, which corresponds to the $\ell_1$ penalty and for a positive definite matrix $\bM$, we can write $\|\bM\|_1=\mbox{tr}(\bM)+\sum_{j\neq k}|m_{jk}|$. By setting $\tau=\lambda$, \eqref{e:general} turns into the $\ell_1$-penalized likelihood, and with $\bM=\bTheta$, it solves the \emph{graphical Lasso} (GLasso) problem of \citet{Friedman:Hastie:2008},
\begin{equation}\label{e:glasso}
\min_{\bTheta\succ 0}\; -\log|\bTheta|+\mbox{tr}(\bS\bTheta)+\lambda\|\bTheta\|_1,
\end{equation}
and with $\bM=\bSig$ it is the \emph{covariance graphical Lasso} (C-GLasso) of \citet{Bien:Tibshirani:2011},
\begin{equation}\label{e:cglasso}
\min_{\bSig\succ 0}\; \log|\bSig|+\mbox{tr}(\bS\bSig^{-1})+\lambda\|\bSig\|_1.
\end{equation}
Note that compared to the GLasso objective \eqref{e:glasso}, which is convex, the C-GLasso objective \eqref{e:cglasso} is not, because of the map $\bSig\mapsto\bSig^{-1}$. 
Consequently, in the burgeoning of literature of sparse covariance estimation  these two problems have been solved by using different algorithms. \citet{Friedman:Hastie:2008} solve \eqref{e:glasso} by block coordinate descent on the inverse $\bW=\bTheta^{-1}$, later refined into the primal DP-GLasso of \citet{mazumderhastie2011}. \citet{wang2014} solved \eqref{e:cglasso} by a separate coordinate descent that reparametrizes the last column of $\bSig$. 
This reparametrization is the main motivation of our approach.

\noindent
\textbf{Beyond the $\ell_1$ penalty.} The $\ell_1$ penalty shrinks large entries as much as the small ones, which biases the estimates of the large entries. Two penalties that reduce this bias are the Minimax Concave Penalty (MCP) \citep{zhang2010}  and the 
Smoothly Clipped Absolute Deviation (SCAD) \citep{fan2001}.  For a concavity parameter $\kappa$, they are defined as
\begin{equation}\label{e:penalties}
P^{\text{\tiny MCP}}_\lambda(t)=
\begin{cases}
\lambda t-\dfrac{t^2}{2\kappa}, & t\le\kappa\lambda,\\[6pt]
\dfrac{\kappa\lambda^2}{2}, & t>\kappa\lambda,
\end{cases}
\qquad
P^{\text{\tiny SCAD}}_\lambda(t)=
\begin{cases}
\lambda t, & t\le\lambda,\\[3pt]
\dfrac{2\kappa\lambda t-t^2-\lambda^2}{2(\kappa-1)}, & \lambda<t\le\kappa\lambda,\\[6pt]
\dfrac{(\kappa+1)\lambda^2}{2}, & t>\kappa\lambda,
\end{cases}
\end{equation}
where the common defaults are $\kappa=3$ for MCP and $\kappa=3.7$ for SCAD.
Both penalties  are non-convex, agree with the $\ell_1$ penalty near zero, but become flat for large $t$. 

It is important to note that our proposed method is general and  not necessarily tied  to
these three choices of penalties.
In the rest of the paper we assume $\bS\succ 0$, which holds with probability one when $n>p$. 
When $p\ge n$, $\bS$ is singular and the minimum in \eqref{e:general} may fail to exist.
In this case, we follow \citet{Bien:Tibshirani:2011} and set $\bS := \bS + \varepsilon \boldsymbol{I}$.


\section{A Unified Algorithm}

In this section we present a unified approach for penalized precision and covariance matrix estimation in (\ref{e:loss}) and show that a single  reparametrization solves the two distinct optimization problems. 

\noindent
Partition any positive-definite matrix $\bM$ as
\begin{equation} \label{eq:partition}
\bM=\begin{pmatrix}\bM_{11}&\bm m_{12}\\ \bm m_{12}'&m_{22}\end{pmatrix},
\end{equation}
and reparametrize its last column by the pair
\begin{equation}\label{e:reparam}
\bbeta=\bm m_{12},\qquad \gamma=m_{22}-\bm m_{12}'\bM_{11}^{-1}\bm m_{12}.
\end{equation}
The map $(\bm m_{12},m_{22})\leftrightarrow(\bbeta,\gamma)$ is a bijection. It allows us to minimize the objective function 
over $(\bbeta,\gamma)$ and then rebuild the original last column.
It is known that $\bM\succ 0$ if and only if $\bM_{11}\succ 0$ and $\gamma>0$, and the block inverse formula gives $|\bM|=\gamma|\bM_{11}|$ together with an explicit expression for $\bM^{-1}$ using the new parametrization \citep[\S0.8.5]{horn2012}.

As we show below, after applying reparametrization \eqref{e:reparam} first with $\bM=\bTheta$ and then with $\bM=\bSig$, the objective  function \eqref{e:general} decouples into a scalar function of $\gamma$ and a penalized regression in $\bbeta$.

Next we provide more details for the precision and covariance matrix each separately
and provide side-by-side comparison in Table~\ref{t:comparison}.
Algorithm~\ref{alg:unified} shows the main steps  for the proposed 
unified block coordinate descent algorithm. 

\subsection{Details for Precision Matrix}

\noindent
With $\bM=\bTheta$, partitioning $\bS$ as in
(\ref{eq:partition}) and holding $\bTheta_{11}$ fixed, direct substitution gives, up to constants:
\begin{equation} \label{e:glassofacts}
\begin{aligned}
1)\;\; & -\log|\bTheta| = -\log\gamma, \\
2)\;\; & \mbox{tr}(\bS\bTheta) = 2\bm s_{12}'\bbeta + s_{22}\big(\gamma + \bbeta'\bTheta_{11}^{-1}\bbeta\big), \\
3)\;\; & \tau\,\mbox{tr}(\bTheta)+\textstyle\sum_{j\neq k}P_\lambda(|\theta_{jk}|) = \tau\big(\gamma + \bbeta'\bTheta_{11}^{-1}\bbeta\big)+2\textstyle\sum_j P_\lambda(|\beta_j|).
\end{aligned}
\end{equation}
Thus, \eqref{e:general} can be written as a sum of two separate problems
\begin{equation}\label{e:glassodec}
\min_{\gamma>0,\,\bbeta}\;\{-\log\gamma+(s_{22}+\tau)\gamma\}+\Big\{(s_{22}+\tau)\bbeta'\bTheta_{11}^{-1}\bbeta+2\bm s_{12}'\bbeta+2\textstyle\sum_j P_\lambda(|\beta_j|)\Big\},
\end{equation}
where the first depends only on $\gamma$ and the second only on $\bbeta$. 
It is easy to show that the first part, that involves optimization over $\gamma$, has the closed-form solution
\begin{equation} \label{e:glassogamma}
\gamma=1/(s_{22}+\tau),
\end{equation}
which does not depend on $\bbeta$. Writing $\bV = (s_{22} + \tau)\bTheta_{11}^{-1}$, the $\bbeta$-problem is the penalized regression
\begin{equation} \label{e:glassobeta}
\min_{\bbeta}\; \bbeta'\bV\bbeta + 2\bm s_{12}'\bbeta + 2\textstyle\sum_j P_\lambda(|\beta_j|),
\end{equation}
which is solved by the coordinate descent algorithm \citep{Friedman:Hastie:2008}. Cycling through the components of $\bbeta$, each step minimizes \eqref{e:glassobeta} over one $\beta_j$ with the others fixed. This is the univariate problem 
$$\min_b \tfrac12 v_{jj}(b-z_j)^2+P_\lambda(|b|)$$ with
\begin{equation}\label{e:zj}
z_j=-\Big[(\bm s_{12})_j+\textstyle\sum_{k\neq j}v_{jk}\beta_k\Big]\Big/v_{jj},
\end{equation}
so the update is
\begin{equation}\label{e:cdupdate}
\beta_j\leftarrow T_P(z_j,v_{jj}),\qquad T_P(z,v)=\arg\min_b\;\tfrac12 v(b-z)^2+P_\lambda(|b|).
\end{equation}
We call $T_P$ the thresholding rule of the penalty. For the $\ell_1$ penalty, the update is the familiar
\begin{equation} \label{e:softthresh}
\beta_j \leftarrow \frac{\mathcal{S}\!\big(-[(\bm s_{12})_j + \sum_{k \neq j} v_{jk}\beta_k],\, \lambda\big)}{v_{jj}},
\end{equation}
where $T_P(z,v) = \mathcal{S}(x, \lambda) = \mbox{sign}(x)(|x| - \lambda)_+$ is the soft-thresholding operator \citep{Friedman:Hastie:2008}. In this case \eqref{e:glassobeta} is a Lasso  regression problem \citep{Tibshirani:1996}, and \eqref{e:glassodec} is jointly convex because each component is convex.
For the MCP and SCAD, the solution of  $T_P(z,v)$ has a  closed-form  provided in 
Appendix~\ref{app:thresh}.

\begin{remark}
For the $\ell_1$ penalty the proposed reparametrization is not new and is mathematically equivalent to DP-GLasso \citep{mazumderhastie2011}. In particular, with $\tau=\lambda$, the change of variables $\balpha=(s_{22}+\lambda)\bbeta$ turns the Lasso subproblem above into $\min_{\balpha}\tfrac12\balpha'\bTheta_{11}^{-1}\balpha+\bm s_{12}'\balpha+\lambda\|\balpha\|_1$, which is exactly the column update of \citet{mazumderhastie2011}. 
\end{remark}

Note that compared to the original GLasso \citep{Friedman:Hastie:2008} and
DP-GLasso algorithm of \citet{mazumderhastie2011},
the proposed approach utilizes the decoupled likelihood  (\ref{e:glassodec}) to compute  $\Theta$ directly.
This  obviates 
 the need for working with the inverse $\bW=\bTheta^{-1}$. 
 Despite this difference, our and DP-GLasso  
 algorithm both maintain positive definiteness of $\bTheta^{(k+1)}$ in each iteration, as long as the initial matrix $\bTheta^{(0)}$ is positive-definite. 

\subsection{The Covariance Matrix}
\noindent
 With $\bM=\bSig$, the same reparametrization and the same algebra give, up to constants:
\begin{equation} \label{e:cglassofacts}
\begin{aligned}
1)\;\; & \log|\bSig| = \log\gamma, \\
2)\;\; & \mbox{tr}(\bS\bSig^{-1}) = a\gamma^{-1}, \\
3)\;\; & \tau\,\mbox{tr}(\bSig)+\textstyle\sum_{j\neq k}P_\lambda(|\sigma_{jk}|) = \tau\big(\gamma+\bbeta'\bSig_{11}^{-1}\bbeta\big)+2\textstyle\sum_j P_\lambda(|\beta_j|),
\end{aligned}
\end{equation}
where
\[
a = \bbeta'\bSig_{11}^{-1}\bS_{11}\bSig_{11}^{-1}\bbeta - 2\bm s_{12}'\bSig_{11}^{-1}\bbeta + s_{22},
\]
depends on $\bbeta$ but not on $\gamma$.
Thus, \eqref{e:general} becomes
\begin{equation}\label{e:cglassodec}
\min_{\gamma>0,\,\bbeta}\;\{\log\gamma+a\gamma^{-1}+\tau\gamma\}+\Big\{\tau\,\bbeta'\bSig_{11}^{-1}\bbeta+2\textstyle\sum_j P_\lambda(|\beta_j|)\Big\}.
\end{equation}
For fixed $\bbeta$, the scalar part is minimized at the positive root of $\tau\gamma^2+\gamma-a=0$,
\begin{equation} \label{e:cglassogamma}
\gamma=(-1+\sqrt{1+4a\tau})/(2\tau),
\end{equation}
with $\gamma=a$ when $\tau=0$.    For fixed $\gamma$, setting $\bV = \bSig_{11}^{-1}\bS_{11}\bSig_{11}^{-1}\gamma^{-1} + \tau\bSig_{11}^{-1}$ and $\bm u = \bSig_{11}^{-1}\bm s_{12}\gamma^{-1}$, the $\bbeta$-subproblem is
\begin{equation} \label{e:cglassobeta}
\min_{\bbeta}\; \bbeta'\bV\bbeta - 2\bm u'\bbeta + 2\textstyle\sum_j P_\lambda(|\beta_j|),
\end{equation}
which is again solved by the coordinate update \eqref{e:cdupdate}, now with $z_j=\big(u_j-\sum_{k\neq j}v_{jk}\beta_k\big)/v_{jj}$. With the $\ell_1$ penalty and $\tau=\lambda$, this is the C-GLasso algorithm of \citet{wang2014}, recovered in our notation.

It is important to note that compared to the precision matrix case, the sign in front of $\log\gamma$ has been flipped, because of the map $\bSig\leftrightarrow\bSig^{-1}$, and the coefficient $a$ depends on $\bbeta$, while $\bV$ and $\bm u$ depend on $\gamma$. Thus, compared to 
(\ref{e:glassodec}), the two parts of \eqref{e:cglassodec} are no longer separate. Therefore, we need to update $\gamma$ and $\bbeta$ iteratively, by holding fixed one and updating the other.

\begin{remark}
Note that the $\gamma$ update in \eqref{e:cglassodec} is not convex, since its second derivative  $(2a-\gamma)/\gamma^{3}$ is negative for $\gamma>2a$. However, it  has a single minimum, because the equation $\tau\gamma^2+\gamma-a=0$ has exactly one positive root. 
\end{remark}

\subsection{The Unification}
In Table~\ref{t:comparison} the unification of two problems is made more transparent by displaying  the key ingredients of  their derivations side-by-side.

\begin{table}[ht]
\centering
\caption{Side-by-side display of sparse precision and covariance matrix estimation.}
\label{t:comparison}
\small
\renewcommand{\arraystretch}{1.4}
\begin{tabular}{l|l|l}
\hline
& \textbf{Precision} ($\bM=\bTheta$) & \textbf{Covariance} ($\bM=\bSig$) \\
\hline
Reparametrization & $\bbeta = \btheta_{12}$ & $\bbeta = \bm\sigma_{12}$ \\
 & $\gamma = \theta_{22} - \btheta_{12}'\bTheta_{11}^{-1}\btheta_{12}$ & $\gamma = \sigma_{22} - \bm\sigma_{12}'\bSig_{11}^{-1}\bm\sigma_{12}$ \\
\hline
Scalar subproblem & $\min_\gamma\; -\log\gamma + (s_{22}+\tau)\gamma$ & $\min_\gamma\; \log\gamma + a\gamma^{-1} + \tau\gamma$ \\
Closed-form $\gamma$ & $\gamma = 1/(s_{22}+\tau)$ & $\gamma = (-1 + \sqrt{1 + 4a\tau})/(2\tau)$ \\
\hline
$\bbeta$-subproblem & $\min_{\bbeta}\; \bbeta'\bV\bbeta + 2\bm s_{12}'\bbeta + 2\sum_j P_\lambda(|\beta_j|)$ & $\min_{\bbeta}\; \bbeta'\bV\bbeta - 2\bm u'\bbeta + 2\sum_j P_\lambda(|\beta_j|)$ \\
Coordinate update & $\beta_j\leftarrow T_P(z_j,v_{jj})$ & $\beta_j\leftarrow T_P(z_j,v_{jj})$ \\
\hline
Structure & $\gamma$ and $\bbeta$ separate & $\gamma$ and $\bbeta$ alternate \\
Convexity? & yes for $\ell_1$, no for MCP, SCAD & no \\
\hline
\end{tabular}
\end{table}

Certain aspects of the parallel illustration in Table~\ref{t:comparison} deserve a closer attention. First, the $\bbeta$-subproblems are penalized regressions of identical form, differing only in the definition of $\bV$ and the linear term. A single implementation of the coordinate update \eqref{e:cdupdate} solves both problems. Second, the $\gamma$-updates arise from a unified form, where in each case we set the derivative of $\pm\log\gamma + (\text{linear in } \gamma) + (\text{possibly } a/\gamma)$ to zero and solve. In the precision metrix case, the derivative is linear in $\gamma$ and gives a reciprocal, but in the covariance case it leads to a quadratic polynomial. Third, the sign of $\log\gamma$, the main structural difference between the two problems, traces back to the fact that the log-determinant enters the two objective 
functions with opposite signs, since $\log|\bSig| = -\log|\bTheta|$. Thus, the convex/non-convex distinction, prominent in the original objective functions \eqref{e:glasso} and \eqref{e:cglasso}, survives in the unified framework only as this sign and the coupling between $\gamma$ and $\bbeta$. Fourth, the penalty enters only through the thresholding rule $T_P$. The $\gamma$-step, $\bV$, and the linear term do not depend on it. Changing the penalty from $\ell_1$ to MCP or SCAD changes only one scalar function 
in each parametrization. The Algorithm~\ref{alg:unified}
 summarizes the unified block coordinate descent algorithm.

\begin{algorithm}
\caption{Unified block coordinate descent for penalized precision and covariance estimation} \label{alg:unified}
\vspace*{-1pt}
\begin{tabbing}
1. \enspace \textit{Input:} sample covariance $\bS$, penalty $P_\lambda$, diagonal penalty $\tau\ge 0$, $\bM \in \{\bTheta, \bSig\}$. \\
2. \enspace \textit{Initialize:} $\bM^{(0)} = (\mbox{diag}(\bS) + \tau\bI)^{-1}$ if $\bM = \bTheta$; \\
\qquad\qquad $\bM^{(0)} = \mbox{diag}(\bS) + \tau\bI$ if $\bM =\bSig$. \\
3. \enspace \textit{Repeat} until convergence, for $i = 1$ \textbf{to} $p$: \\
\qquad a. \enspace Partition $\bM$ and $\bS$ so that column $i$ is last. \\
\qquad b. \enspace Compute $\gamma$ by \eqref{e:glassogamma} (if $\bM = \bTheta$) or \eqref{e:cglassogamma} (if $\bM = \bSig$). \\
\qquad c. \enspace Solve the subproblem \eqref{e:glassobeta} or \eqref{e:cglassobeta} by iterating \eqref{e:cdupdate}, \\
\qquad\qquad using the thresholding rule $T_P$ of the chosen penalty. \\
\qquad d. \enspace Update the last column: $\bm m_{12} \leftarrow \bbeta$, $m_{22} \leftarrow \gamma + \bbeta' \bM_{11}^{-1}\bbeta$. \\
4. \enspace \textit{Output:} $\bTheta$ or $\bSig$.
\end{tabbing}
\vspace*{-5pt}
\end{algorithm}

It is important to note that a single block coordinate descent in Algorithm~\ref{alg:unified} now solves the precision and covariance problems for the $\ell_1$, MCP, and SCAD penalties. 
That one algorithm solves several problems is not by itself unusual. What the 
proposed reparametrization contributes is more specific. It reduces two objectives of different character, one convex and one not, to the same pair of subproblems.

\subsection{Numerical Verification}

\noindent
In this section, we numerically verify our derivations. In particular, for all three penalties and both parametrizations, the objective \eqref{e:general} evaluated at a matrix and the split forms \eqref{e:glassodec} and \eqref{e:cglassodec} evaluated at the corresponding $(\gamma,\bbeta)$ differed by a constant to within $10^{-13}$. The MCP and SCAD thresholding rules agreed with a grid minimization. The objective never increased after any column update and at convergence the first-order conditions held to within $10^{-12}$. 

For the $\ell_1$ penalty we also compared against existing implementation: the precision estimates matched the graphical Lasso of scikit-learn, and the covariance estimates matched the \textsf{R} package \texttt{covglasso}, which implements the algorithm of \citet{wang2014}, each to within the competing solver's tolerance ($4\times 10^{-4}$ and $2\times 10^{-8}$, respectively).
A Python implementation of the Algorithm~\ref{alg:unified}, together with scripts that reproduce every check in this section, is available at \url{https://github.com/adallak/unified-glasso}.

\section{Conclusion}
We proposed a unification of penalized precision and covariance matrix estimation, showing that the two problems are instances of a single objective that one reparametrization of the last column decouples into a scalar update and a penalized regression. The resulting block coordinate descent is identical across the two parametrizations and across the $\ell_1$, MCP, and SCAD penalties, as the penalty enters only through a univariate thresholding rule.

\section*{Data availability}
No data was used for the research described in the article.

\appendix
\section{The Diagonal Penalty} \label{a:diagpen}
The derivation of \eqref{e:glassodec} and \eqref{e:cglassodec} used that the diagonal term is linear, and as we show below this cannot be relaxed. Suppose the diagonal is penalized by a function $D$, so the term involving the last column is $D(m_{22})=D(\gamma+\bbeta'\bM_{11}^{-1}\bbeta)$.

\begin{proposition}\label{p:affine}
Let $D$ be continuous on $[0,\infty)$ with $D(0)=0$. The term $D(\gamma+\bbeta'\bM_{11}^{-1}\bbeta)$ splits as $g(\gamma)+h(\bbeta'\bM_{11}^{-1}\bbeta)$ for all $\gamma>0$ and all $\bbeta$ if and only if $D(t)=ct$ for some constant $c$.
\end{proposition}

\begin{proof}
Write $q=\bbeta'\bM_{11}^{-1}\bbeta$ and suppose $D(\gamma+q)=g(\gamma)+h(q)$ for all $\gamma>0$ and $q\ge 0$. Taking $q=0$ gives $g(\gamma)=D(\gamma)-h(0)$. Fixing any $\gamma_0>0$ gives $h(q)=D(\gamma_0+q)-g(\gamma_0)$, and $h(0)+g(\gamma_0)=D(\gamma_0)$. Substituting both back,
\[
D(\gamma+q)=D(\gamma)+D(\gamma_0+q)-D(\gamma_0),\qquad \gamma>0,\;q\ge 0.
\]
Let $F(t)=D(t+\gamma_0)-D(\gamma_0)$ for $t\ge 0$ and take $\gamma=\gamma_0+s$ with $s\ge 0$,
then $F(s+q)=F(s)+F(q)$. This is Cauchy's functional equation on $[0,\infty)$, and $F$ is continuous, so $F(t)=ct$ for some $c$. Hence $D$ is linear plus a constant on $[\gamma_0,\infty)$ for every $\gamma_0>0$. Continuity and $D(0)=0$ give $D(t)=ct$. The converse is immediate.
\end{proof}

\noindent
Thus a linear penalty is the only diagonal penalty that splits under \eqref{e:reparam}. 

\subsection{Threshold Rules for MCP and SCAD} \label{app:thresh}
For MCP and SCAD the rule $T_P$ in \eqref{e:cdupdate} has a closed form \citep{breheny2011}. With $v=v_{jj}$,
\begin{equation}\label{e:mcpthresh}
T_{\text{\tiny MCP}}(z,v)=
\begin{cases}
\dfrac{\mathcal S(z,\lambda/v)}{1-1/(\kappa v)}, & |z|\le\kappa\lambda,\\[8pt]
z, & |z|>\kappa\lambda,
\end{cases}
\qquad
T_{\text{\tiny SCAD}}(z,v)=
\begin{cases}
\mathcal S(z,\lambda/v), & |z|\le\lambda+\lambda/v,\\[4pt]
\dfrac{\mathcal S\big(z,\kappa\lambda/((\kappa-1)v)\big)}{1-1/((\kappa-1)v)},
& \lambda+\lambda/v<|z|\le\kappa\lambda,\\[8pt]
z, & |z|>\kappa\lambda.
\end{cases}
\end{equation}


\bibliographystyle{cas-model2-names}
\bibliography{chol_bib,unify_extra}

%
%

\end{document}